*Letter*

# Spatiotemporal Patterns of COVID-19 Impact on Human Activities and Environment in China Using Nighttime Light and Air Quality Data

**Qian Liu**[1,2], **Dexuan Sha**[1,2], **Wei Liu**[1,3], **Paul Houser**[2], **Luyao Zhang**[4], **Ruizhi Hou**[5], **Hai Lan**[1,6], **Colin Flynn**[2], **Mingyue Lu**[7], **Tao Hu**[8] **and Chaowei Yang** [1,*]

1. NSF Spatiotemporal Innovation Center, George Mason Univ., Fairfax, VA 22030, USA; qliu6@gmu.edu (Q.L.); dsha@gmu.edu (D.S.)
2. Department of Geography and Geoinformation Science, George Mason Univ., Fairfax, VA, 22030, USA; phouser@gmu.edu (P.H.); cflynn8@gmu.edu (C.F.)
3. College of Land Science and Technology, China Agricultural University, Beijing, 100083, China；devilweil@cau.edu.cn (W.L.);
4. School of Business Management, East China Normal University, Shanghai, 3663 Zhongshanbei Road, China; sunshineluyao@gmail.com; (L.Z.)
5. School of Mathematical Sciences, East China Normal University, Shanghai, 3663 Zhongshanbei Road, China; houruizhiecnu@gmail.com; (R.H.)
6. Department of Geographical Sciences, University of Maryland, College Park, MD 20742, USA; hlan@terpmail.umd.edu (H.L.)
7. School of Geographical Sciences, Nanjing University of Information Science & Technology, 219 Ningliu Road, Nanjing, Jiangsu 210044, China; lumingyue@nuist.edu.cn (M.L.)
8. Center for Geographic Analysis, Harvard University, Cambridge, MA 02138, USA; taohu@g.harvard.edu (T.H.);
* Correspondence: cyang3@gmu.edu (C.Y.)

**Abstract:** In order to analyze the impact of COVID-19 on people's lives, activities and the natural environment, this paper investigates the spatial and temporal characteristics of Night Time Light (NTL) radiance and Air Quality Index (AQI) before and during the pandemic in mainland China. Our results show that the monthly average NTL brightness is much lower during the quarantine period than before. This study categorizes NTL into three classes: residential area, transportation and public facilities and commercial centers, with NTL radiance ranges of 5~20, 20~40 and greater than 40 ($nW \cdot cm^{-2} \cdot sr^{-1}$) [1], respectively. We found that the Number Of Pixels (NOP) with NTL detection increased in the residential area and decreased in the commercial centers for most of the provinces after the shutdown, while transportation and public facilities generally stayed the same. More specifically, we examined these factors in Wuhan, where the first confirmed cases were reported, and where the earliest quarantine measures were taken. Observations and analysis of pixels associated with commercial centers were observed to have lower NTL radiance values, indicating a dimming behavior, while residential area pixels recorded increased levels of brightness, after the beginning of the lockdown. The study also discovered a significant decreasing trend in the daily average AQI for the whole country, with cleaner air in most provinces during February and March, compared to January 2020. In conclusion, the outbreak and spread of COVID-19 has had a crucial impact on people's daily lives and activity ranges through the increased implementation of lockdown and quarantine policies. On the other hand, the air quality of China has improved with the reduction of non-essential industries and motor vehicle usage.

## 1. Introduction

At the end of 2019, the sudden outbreak of the coronavirus disease (COVID-19) in Wuhan City, China, required mitigation and containment measures that brought the most populous country and second largest economy in the world to a halt. Coincident with the Chinese spring festival migration, cases quickly spread across the whole country. Despite China's effective efforts in containing the outbreak since late January, within about three months, COVID-19 has been found in 185 countries, and taken more than 162 thousand lives. Worldwide, more than two million cases have been confirmed by the second half of April 2020, and the number is still growing rapidly. Wuhan City was locked down on Jan. 23, 2020, as soon as the outbreak and contagious nature of the novel coronavirus was confirmed. In ten days, most of the provinces in Mainland China implemented strict policies to limit non-essential activities, transportation and production. Hundreds of millions of people volunteered to stay at home and practice self-isolation. Consequently, drastic changes suddenly took place in everyone's lives, influencing their work and living environments.

Currently, with the recent reopening of Wuhan, the spread of COVID-19 is been considered to be initially controlled in China. However, the pandemic situation is still severe and grim in the other parts of the world; there are more than 700,000 confirmed cases in the U.S., as of April 18, 2020. A detailed and quantitative analysis of the impact of COVID-19 has had on people's lives, activities and the environment, as well as the citizens' response to the policies enacted to control the disease in China is urgent and significantly important for the reference of other countries. Studies and analytics have been conducted to estimate, assess and predict the spread of this pandemic using temperature [2,3], humidity [4,5], air quality [6], geometry [7] and migration data [8], and how the control measures impact the outbreak [9]. But very few published articles in the scientific literature illustrate the impact of virus mitigation and containment measures on the environment. Our study aims to fill this gap because understanding the impact of COVID-19 is important for decision making in regards to post-pandemic reopening strategies, economic loss assessments and environment regulations.

Remote sensing data and technologies are widely used in the study of natural disasters and the spread of epidemics [10, 11, 12, 13]. Night Time Light (NTL) images have the capability to not only tell people where the facilities are, but also depict when and how they are used. Therefore, the citizens' reaction to the lockdown and quarantine policies can be reflected and monitored. NASA has released some comparisons between the NTL imagery before and during the lockdown of Wuhan [14], and has demonstrated that some districts and highways are visually dimming, attributed to the COVID-19 shutdown. However, there are no quantitative statistics regarding exactly which area are less bright. Some questions remain to be answered, such as: Are all the parts of the city getting dimmer? Are the conditions similar in other provinces? Could some areas stay the same or even get brighter at night? Answers to these questions are crucial in understanding the changes in policy-driven patterns caused by the pandemic, which in turn is important for analyzing the influence of COVID-19 and decision making on the response methods. To answer these questions, this study first calculates and compares the monthly average NTL radiance values before and during the pandemic for each province, then categorizes them into three classes: (1) residential areas, (2) transportation facilities and urban infrastructures, and (3) commercial centers, and finally analyzes the change of Number Of Pixels (NOP) in each category. Spatiotemporal analytics on average NTL is also conducted to achieve a better comprehension of the impact. Through these categorizations and investigations, the influence of COVID-19 on human activities can be efficiently depicted.

Air quality is believed to have a robust interaction with COVID-19. It was reported to influence the mortality outcome of COVID-19 [15], researchers found that an increase of only 1 μg/$m^3$ in $PM_{2.5}$ is associated with a 15% increase in the death rate. [6] studied the association between meteorological factors and COVID-19 transmission and illustrated that bad air quality with strong wind will accelerate the spread of the virus. On the other hand, the pandemic situation also impacts the air quality indirectly. Both NASA and the European Space Agency (ESA) have observed plummets in airborne Nitrogen Dioxide ($NO_2$) over China using modified Copernicus Sentinel 5P data [16, 17]. Moreover, [18] illustrated that the outbreak of COVID-19 has forced China to enact lockdown procedures, suppressing its industrial activities and, hence, dropped its $NO_2$ and carbon emissions by 30% and 25%, respectively. All of these discoveries focused on the change of one or two pollutants induced by COVID-19, but the total air quality cannot be determined without the consideration of other air pollutants, such as Sulfur Dioxide ($SO_2$), Carbon Monoxide (CO), Ozone and particulate matters (PMs). Our study investigates the ground-based air quality index (AQI), which has been derived from the concentration of five major air pollutants before and during the COVID-19 crisis for every province in mainland China and offers a better vision of the COVID-19 impact on the overall air quality.

**2. Materials and Methods**

*2.1 Dataset*

The daily NTL data used in the study are the VNP46A1 product derived from the Visible Infrared Imaging Radiometer Suite (VIIRS) sensor Day Night Band (DNB) onboard the Suomi National Polar-orbiting Partnership (NPP) Satellite and archived at NASA's LAADS DAAC data center (https://ladsweb.modaps.eosdis.nasa.gov/), with a study period from Dec. 1, 2019 to Mar. 26, 2020. These data have been corrected for clouds, atmospheric conditions, terrain, vegetation, snow, lunar illumination, and stray light effects, using the Black Marble algorithm [19] and gridded to 500 meters globally. The VIIRS/DNB has substantial improvements in sensor resolution and calibration to guarantee fewer over-glow and saturation effects when compared to the previous era of Defense Meteorological Satellite Program/Operational Linescan System's (DMSP/OLS) nighttime lights image products. The VIIRS/DNB has a spectral range of 500–900 nm that is highly sensitive to very low levels of visible light and can significantly improve the ability to detect anthropogenic lighting from buildings, roads, and other city infrastructures at night without moonlight influence [20]. The cloud masks associated with NTL imagery are performed using the VIIRS thermal band (band M15) with a spectral range of 10.26–11.26 μm [21]. Although there are NTL data with higher spatial resolutions such as Luojia1-01, they are not fully accessible for international users.

The AQI is collected from the China National Environmental Monitoring Center (CNEMC) (http://www.cnemc.cn/) with a period from Jan. 1, 2020 to Mar. 28, 2020. These data are calculated from the concentration of air pollutants such as $PM_{2.5}$, $PM_{10}$, Ozone ($O_3$), Nitrogen Dioxide ($NO_2$), Sulfur Dioxide ($SO_2$) and Carbon Monoxide (CO) [22] and used to measure the severity of the air pollution.

*2.2 Data preprocessing*

2.2.1 NPP/VIIRS NTL data

VIIRS NTL data are preprocessed as follows:
1. All the granules of DNB images covering Mainland China are merged with longitude, latitude and cloud mask information;
2. Clouds impact the NTL by reducing the intensity of the lights and blurring the spatial detail of the light features [23]; as a result, the "confidentially" and "probably" cloudy pixels are marked as invalid according to the cloud mask.
3. Monthly nighttime radiance maps are obtained by averaging cloud-free pixels' brightness of each grid in every month during the study periods, which are December of 2019, January, February and March of 2020.
4. The NTL maps derived from step 3 are rectified with the Chinese provincial using a base map.
5. The statistical parameters, such as monthly average radiance and the NOP of each NTL category are calculated for every province and the whole country. Our study focuses on the human activities and artificial lights that mainly exist in residential, urban and built-up areas. The pixels with radiance values lower than $5 \text{ nW} \cdot cm^{-2} \cdot sr^{-1}$ are excluded, which is the spectral range for vegetation, water, snow and rural areas [24].

2.2.2 AQI data

The preparation of AQI data includes the following steps:
1. Rectify meteorological stations with the provincial base map.
2. Calculate the daily and monthly average AQI for each station from the original hourly data.
3. Calculate the mean value of all the stations in each province and the whole country.

*2.3 NTL radiance categorization*

COVID-19 crisis mitigation and containment require people to reduce their outdoor activities, crowding and traveling. In order to analyze these changes on human activities, this paper categorizes the NTL radiances into three classes according to the brightness values, effectively reflecting the different anthropogenic functional zones, including residential areas, transportation and public facilities, and commercial centers. Previous research adopted NTL radiance thresholds to identify these regions [25, 26, 27, 28]. [1] investigated the pixel-level relationship between the VIIRS DNB-derived NTL, Landsat-derived land-use/land-cover, and a map of point of interest (POI) density over China, especially the artificial surfaces in urban land. According to their results, 5~20, 20~40 and greater than 40 ($nW \cdot cm^{-2} \cdot sr^{-1}$) are appropriate radiance ranges for residential areas, transportation and public facilities and commercial centers, respectively.

**3. Results**

*3.1 Impact of COVID-19 on human activity observed by NTL*

Figure 1 shows the average NTL radiance for every province and all of China. During January and February 2020, COVID-19 broke out across China, and most provinces implemented strict quarantine policies, resulting in the majority of people spending their time at home. The average artificial NTL radiance levels during this period are much lower than December 2019 in most provinces, except for Inner Mongolia, Xinjiang, Xizang and Qinghai, where the crisis is not severe, with 194, 76, 1 and 18 cases respectively. Additionally, the NTL of these provinces prior to the outbreak is not bright compared to the other provinces. With the reopening of most provinces in March 2020, the average NTL brightness has increased in comparison to January and February. However, exceptions to this have been observed in Hubei province, with its capital city Wuhan being the center of the outbreak, and several other provinces near it that contain larger numbers of confirmed cases, such as Hunan (1019 cases), Jiangxi (937 cases) and Guangdong (1585 cases). For the national level (chart in the lower right corner), a decreasing trend can be found from December 2019 to February 2020. Although the average NTL brightness almost recovered in March, the value is still a little lower than December's. The reopening was carried out in steps and hadn't been totally completed by the end of March in some of the provinces mentioned above, influencing the average of NTL for the whole country.

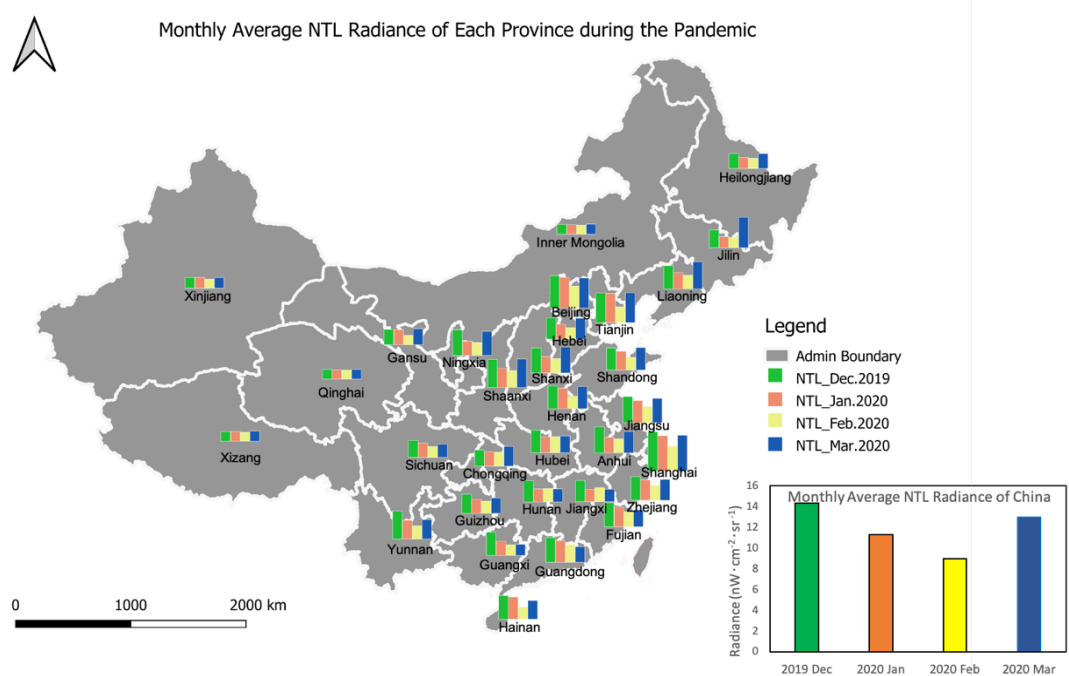

**Figure 1.** Monthly average NTL for each province and China (lower right corner) from Dec. 2019 to Mar. 2020

Figure 2 shows the differences of NOP in the three NTL categories between the first three months (January, February and March) of 2020 and December 2019 (the former minus latter), in each province and all of China. Because of the quarantine policies conducted in most provinces, people stopped going shopping, getting together and working in the office which mostly takes place in commercial centers, or so-called Central Business Districts (CBD) at night. Instead, most people stayed at home, in the residential areas. These changes in human activity are reflected in the statistical results of the NTL categorization:

For the 5~20 nW·$cm^{-2}·sr^{-1}$ category, indicative of residential areas, 20 provinces have higher NOP in all three of the studied months of 2020 than in December 2019. Only Xingjiang in January, and Chongqing, Heilongjiang, Jilin, Liaoning and Xinjiang in March have decreasing numbers larger than 10,000. All the negative differences are smaller than 10,000 in February, when the crisis was most rigorous. These spatiotemporal patterns are due to the stay-at-home policies and more people spending their nights at home.

For the 20~40 nW·$cm^{-2}·sr^{-1}$ group, which reflects the anthropogenic NTL of transportation facilities and public infrastructures, the numbers increased in January and decreased in February and March for most provinces. The quarantine policies were implemented at the end of January and the transportation and public facilities were busier in the first three weeks than in December due to the approaching Spring Festival. The corresponding areas are therefore brighter in January, on average. Negative values are found in most provinces for February and March because the virus spread rapidly, and the public and transportation infrastructures were not utilized as frequently as usual.

For the commercial-center pixels, with NTL brightness higher than 40 nW·$cm^{-2}·sr^{-1}$, their numbers decreased significantly during all three months, compared to before in most provinces. People tended to avoid going to shopping and entertainment centers due to the quarantine policies and fear of getting infected. Although most cities have reopened and loosened their restrictive policies, people are still cautious to gather in crowds, and most shopping malls are closing earlier in the evenings. There are positive values in Henan, Shaanxi and Tianjin, but the increases are all under 500 pixels which can be due to differences in seasonal variations, sensor's condition or snow fall.

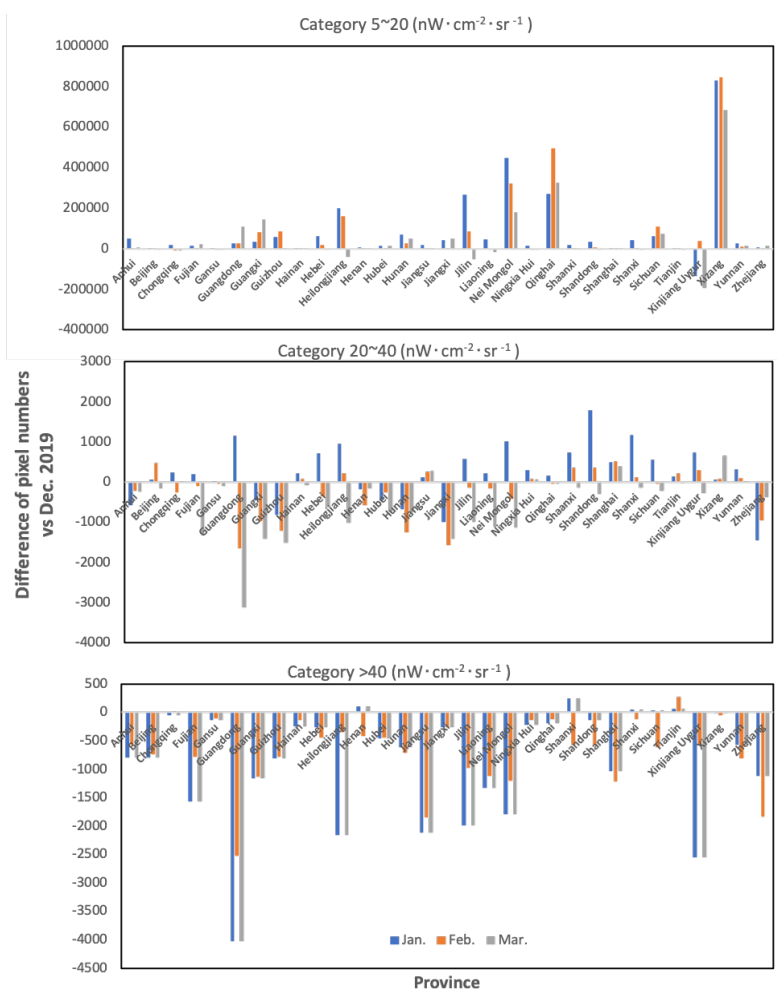

**Figure 2.** Differences of NOP in the three NTL categories between the first three months of 2020 and Dec. 2019, in each province of China

Figure 4 (a) and (b) are the NTL images of Wuhan before and during the lockdown, and (c) is the difference between (b) and (a). The regions in the yellow and green circles are the Jianghan and Guanggu commercial centers, respectively, which are two of the most prosperous and crowded areas in Wuhan. We can observe that these regions are dimmer during the lockdown, in February 2020, compared to December 2019. The differences also show mostly negative values in these two circles. Contrastingly, residential areas outside of the commercial centers are brighter in January, and the differences are obviously positive in those regions. However, the pixels on highways and main roads are not clearly getting brighter or dimmer, they tend to show a mixed texture. This phenomenon can also be verified in the difference map (Figure 3(c)).

Figure 3(d) is the NOP difference in the three categories between the first three months of 2020 and December 2019 (the former minus latter). For all the months during the city lockdown, there are significantly more pixels in the residential category and less pixels in the commercial centers than before. The number of transportation and public facilities lights decreased in January and March while increasing in February. Besides the explanations that have already been discussed, in February, large amounts of food, living and medical resources were transported to Wuhan from all over China and some other countries around the world; national and private medical teams volunteered to support and traveled to Wuhan through highways and airports; temporary hospitals such as Huoshenshan and Leishenshan were built rapidly. Consequently, the NOP in this category is higher during February than December.

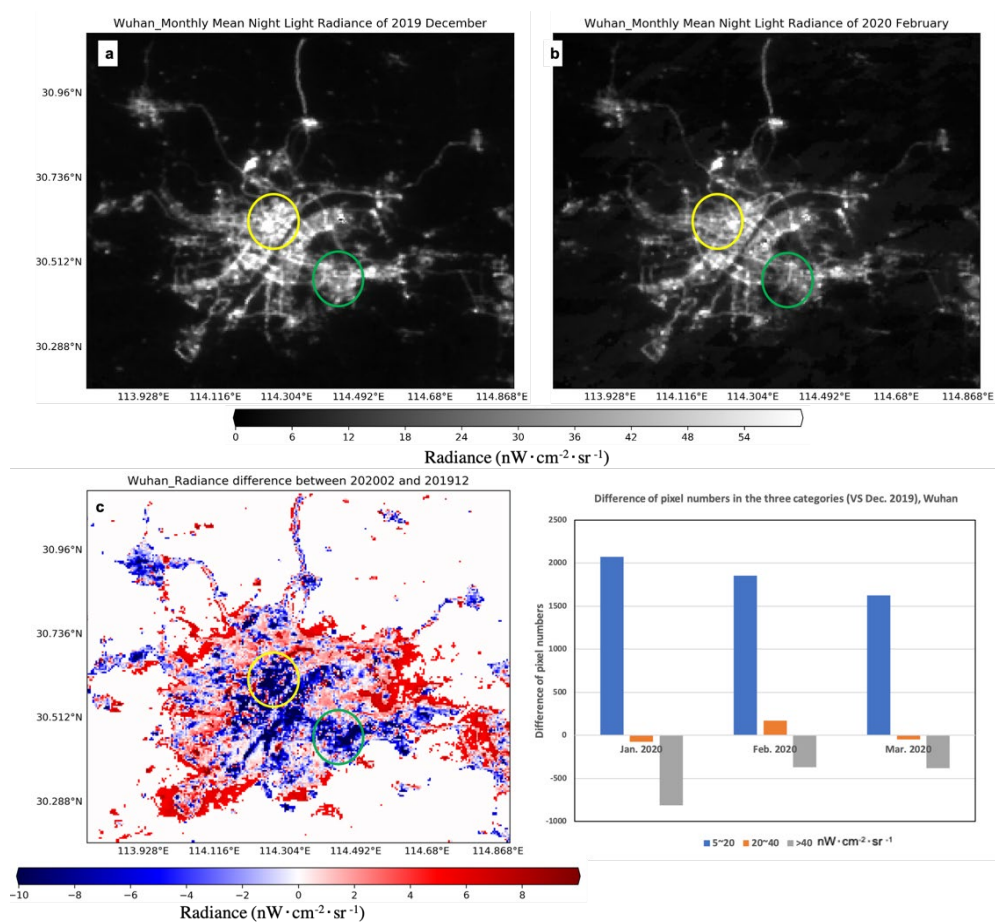

**Figure 3.** (a) Monthly average NTL radiance of Wuhan before lockdown; (b) Monthly average nighttime light radiance of Wuhan after lockdown; (c) Difference between (b) and (c); (d) Differences of NOP in the three NTL categories between first three months of 2020 and Dec. 2019, in Wuhan.

*3.2 Impact of COVID-19 on air quality*

The main air pollution sources in China are power plants, industrial facilities, automobiles, biomass burning, and fossil fuels used in homes and factories for heating [29, 30, 31]. Figure 4 is the daily average AQI time series of China from Jan. 1, 2020 to Mar. 26, 2020. A significant decreasing trend can be seen through the data with a p-value smaller than 0.001. With the breakout of COVID-19, Chinese central and local governments executed strict policies on restricting the production of non-essential industries, traveling by private motor vehicles, which are among the primary sources of air pollutant emissions. As a consequence, the total air quality improved and the AQI decreased in this period. A valley point can be observed on Feb. 15, 2020, nearly a week after most cities in China ordered shutdown policies around Feb. 7~10, 2020. Although the production and travel restrictions orders were applied, the dispersion and deposition of air pollutants is not an instantaneous procedure but needs a period of time depending on the atmospheric conditions, including wind speed and directions, temperature, air pressure, etc. [32], as well as geographic characteristics such as vegetation and mountains [33]. Therefore, the time series shows a time lag of several days for the atmospheric response.

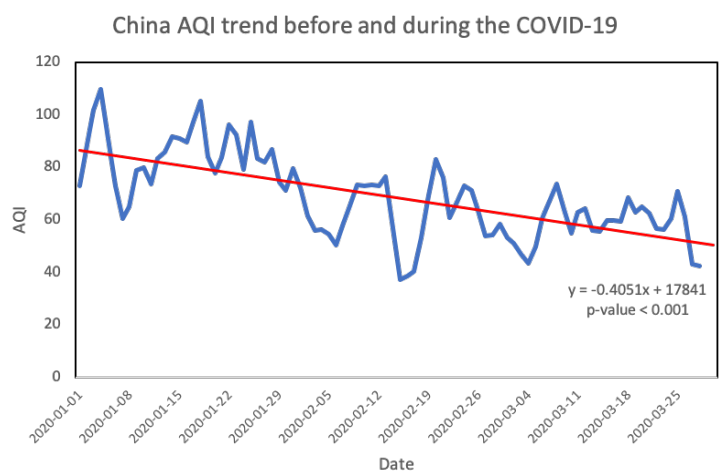

**Figure 4.** Daily average AQI time series of China from Jan. 1, 2020 to Mar. 26, 2020

Figure 5 is the monthly average AQI of each province and the whole Mainland China (lower right corner) from January to March 2020. The air pollution of most provinces is less severe in February and March than January, especially for the Northeast, Northwest, east coast and central China. The reduction of many non-essential industries and private vehicles are the main reasons for the better air quality. Although Wuhan began its lockdown on Jan. 23, 2020, all other cities were still open until the first week of February. The pollutant emissions from the industries and vehicles did not change much during January. Exceptions can be found in Xinjiang, Qinghai, Gansu, Yunnan and Guizhou where the air quality became worse in February and March. For these provinces, the number of confirmed COVID-19 cases are relatively small, totally 78, 18, 139, 184 and 147, respectively, as of Apr. 20, 2020. Therefore, the quarantine policies were not as strict as in other areas and many industries kept working during the study period. Moreover, the winter season, with its dry air conditions, is not conducive for the dispersion of air pollutants [29]. The total air quality condition of China also shows improvement for the three months.

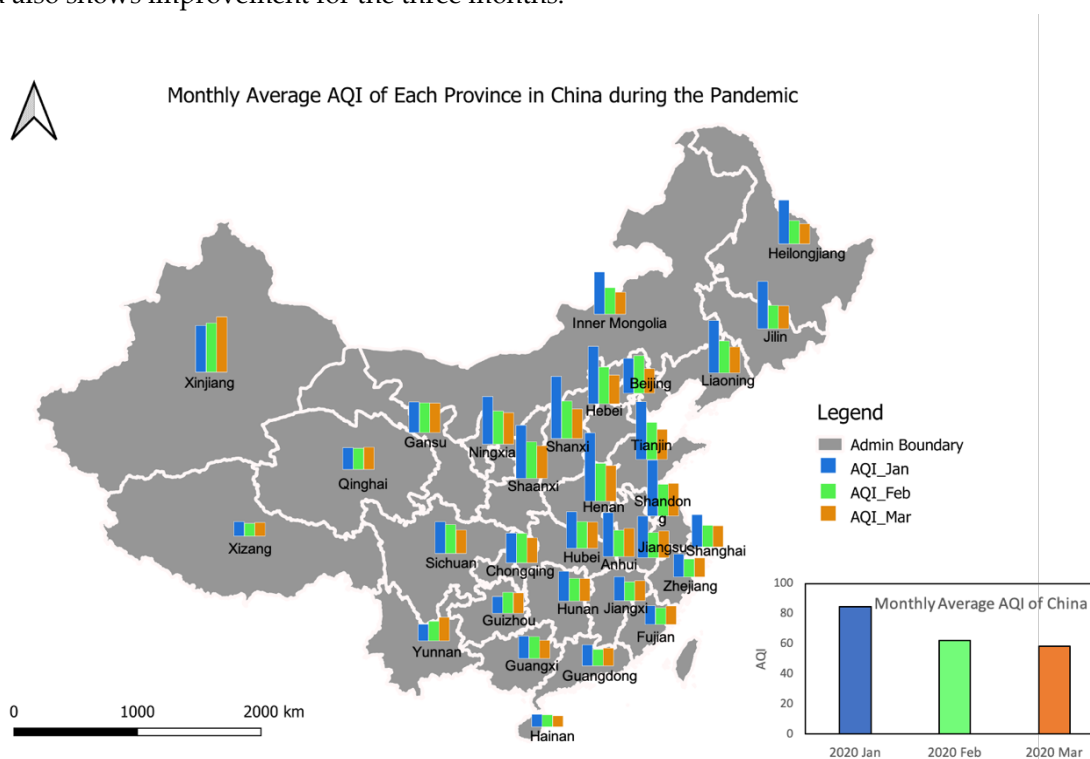

**Figure 5.** Monthly average AQI of each province and China (lower right corner) from Jan. 2020 to Mar. 2020

## 4. Discussion

The impact of the COVID-19 crisis on the atmospheric environment and human activities are studied in this paper by comprehensively investigating the NTL radiance and AQI of every province and the entire Mainland China before and during the pandemic in both spatial and temporal dimensions. The activity ranges of citizens among different functional zones changed dramatically due to the COVID-19 related policies. The CBD and commercial centers became less bright in the NTL images and residential regions lit up. Most people followed the instructions and suggestions of the government and health department strictly to fight against the crisis and protect themselves. Although these changes and impacts bring unprecedented inconvenience to people's daily lives and work, the spread of COVID-19 was initially controlled in about two months. This gives an efficient reference for the rest of the world to make their own policies, strategies and decisions regarding how to control the spread of the virus. Since people stayed at home for nearly two months, some of their consumption and working habits have changed. For example, more people will be accustomed to online purchases and working from home, which will influence economic structures and the incomes of retailers. It has already been proven that NTL has a close relationship with the economy [34, 35]; it should be used in the investigation of post-pandemic socioeconomic situations.

In spite of all the worldwide catastrophes it has brought to human lives, health, economies and medical resources, COVID-19 has made our air cleaner. This may have some positive influences on the recovery of other respiratory diseases. Analytics are worth being done with medical data involved, which will give the decision makers more comprehensive support on the medical and first-aid resources arrangement.

This paper tends to offer short-term qualitative analytics on how people and environment response to the COVID-19 crisis. The seasonal cycles, construction development and Spring Festival influences are not considered quantitatively in the analysis. In the future, all these factors will be calculated and removed from the trends by investigating long-term historic data to conduct quantitative analysis on correlation, regression and prediction.

Despite the effort made by this paper and all the other research, more work is needed to be done on the impact of COVID-19, including:

1. The socioeconomic impact of COVID-19 will be examined by monitoring the change in economic conditions such as GDP, individual income and unemployment rate using NTL and census data.
2. As some analytics and news illustrated, the infection and death rates of COVID-19 have various patterns in different communities [36]. The NTL and other high-resolution remote sensing data sources can be used to distinguish community types in terms of income levels, races and occupations [37]. In the future, the COVID-19 spread and impact condition will be further studied in different human groups.
3. The investigation on air quality will be more detailed on some specific pollutants such as $SO_2$, CO and Ozone that are not addressed in the previous studies.
4. Since the COVID-19 has shown its effects on atmospheric conditions, will it influence the weather, or even the climate, if it cannot be controlled in a short time? Further research is needed on the impact modellings with more climatic and virus-spread factors.

## 5. Conclusions

This research provides detailed spatiotemporal analytics on the impact of COVID-19 on human lives as reflected by nighttime light and air quality. From the results of the experiments and statistics, we conclude:

1. The average NTL radiance decreases in most provinces and the entire country of China with the implementation of shutdown policies. Some exceptions are shown in several provinces due to their small number of confirmed cases, quarantine policies and low original NTL brightness.
2. The number of detected NTL pixels increases in the residential areas while it decreases in the commercial center regions, and generally stays the same in the transportation and public facilities during the studied pandemic time period. This reflects a transfer of human activities from shopping and entertainment centers to residential areas due to the quarantine policies.
3. The total air quality improved during the COVID-19 crisis because of the reduction in industrial production and vehicle usage.
4. The spread of COVID-19 and related policies have a significant impact on people's daily lives and the environment.

These analytics can be further advanced with spatiotemporal and big data methodologies [38] for restraining the spread of COVID-19, medical resources arrangement, planning and implementing of reopening policies, estimating the economic loss and making financial plans.

**Author Contributions:** C.Y., Q.L. and P.H. came up with the original research idea; C.Y. advised Q.L., D.S., and W.L. on the experiment design and paper structure; Q.L. designed the experiments, developed the scripts for experiments, conducted the experiments, and analyzed the experiment results; H.L. supported computing for data processing; D.S. produced the provincial base map, NTL and AQI maps; Q.L. and W.L. produced the NTL results; L. Z., R.H., M.L. and T.H. acquired and cleaned the AQI data. Q. L. wrote the paper; C. Y., P.H. and C.F. revised the paper.

**Funding:** This research was funded by NSF (1841520 and 1835507).

**Acknowledgments:** VIIRS NTL data are obtained from the open accessible NASA GES DISC. China National Environmental Monitoring Center provided the AQI data.

**Conflicts of Interest:** The authors declare no conflict of interest.